\documentclass[journal]{IEEEtran}
\ifCLASSINFOpdf
\else
\fi
 
\usepackage{url}
\usepackage{graphicx}
\usepackage{epstopdf}
\usepackage{float}
\usepackage{color}
\usepackage{amsmath}
\usepackage{array}
\usepackage{relsize}
\usepackage{subfigure}
\usepackage{amssymb,latexsym}
\usepackage{multirow}

\begin{document}
%
\title{ CSSTag: Optical Nanoscale Radar and  Particle Tracking for In-Body and Microfluidic Systems with Vibrating Graphene and Resonance Energy Transfer}

\large\author{
\centering \vspace{.2cm}{Burhan Gulbahar,~\IEEEmembership{Senior Member,~IEEE}} and {Gorkem Memisoglu,~\IEEEmembership{Member,~IEEE}} \\
\thanks{Dr. Burhan Gulbahar is with the Department of Electrical and Electronics Engineering and Applied Research Center of Technology Products, Ozyegin University, Istanbul, 34794, Turkey, (e-mail: burhan.gulbahar@ozyegin.edu.tr). Dr. Gorkem Memisoglu is with the Research Group in Vestel Electronics, Manisa, Turkey, (e-mail: gorkem.memisoglu@vestel.com.tr). }
}
 

\maketitle
 

\begin{abstract}
Single particle tracking  systems  monitor cellular processes with great accuracy in nano-biological systems. The emissions of the fluorescent molecules are detected with cameras or photodetectors. However, state-of-the-art imaging systems have challenges in the detection capability, collection and analysis of imaging data,  penetration depth and complicated set-ups. In this article, a \textit{signaling based nanoscale acousto-optic radar and microfluidic particle tracking system} is proposed based on the theoretical design providing nanoscale optical modulator with vibrating F{\"{o}}rster resonance energy transfer (VFRET) and vibrating CdSe/ZnS quantum dots (QDs) on graphene resonators. The modulator structure combines the significant advantages of graphene membranes having wideband resonance frequencies with QDs having broad absorption spectrum and tunable properties. The solution denoted by chirp spread spectrum (CSS) Tag (\textit{CSSTag})  utilizes classical radar target tracking approaches in nanoscale environments based on the capability to generate CSS sequences to identify different bio-particles. Numerical and Monte-Carlo simulations are realized showing the significant performance for multiple particle tracking (MPT) with a modulator of  $10 \, \mu$m  $\times$ $10 \, \mu$m  $\times$ $10 \, \mu$m  dimension and several picograms of weight, signal to noise ratio (SNR)  in the range $-7$ dB to $10$ dB and  high speed tracking capability for microfluidic and in-body environments.
\end{abstract}



\IEEEpeerreviewmaketitle

\section{Introduction}
\label{introduction}

\IEEEPARstart{S}{ingle} particle tracking (SPT) allows observation of dynamic behaviours of  biological particles with precision better than the diffraction limit of light \cite{spt1}. Special fluorescing tags attached on the molecules are tracked and the digital image processing tools allow nanometer (nm) resolution in positioning. Image processing methods and algorithms are developed in significantly many studies for  particle and cell tracking in microscopic platforms  \cite{im2, im3, im4}. However, the tags do not have signaling capability to support a \textit{signaling based tracking} compared with traditional \textit{imaging based tracking} systems, e.g., fluorescence lifetime imaging for cell tracking with near infrared emissive polymersomes in \cite{flim}. In this article, the existing imaging based particle tracking methods are, for the first time, improved with a special theoretical design of \textit{nanoscale optical radar architecture} denoted by \textit{CSSTag} generating chirp spread spectrum (CSS) modulation by utilizing a recently introduced nanoscale optical modulation mechanism in \cite{bg1, bga, bgb, bgc} defining vibrating F{\"{o}}rster resonance energy transfer (VFRET) of CdSe/ZnS quantum dots (QDs) on graphene membranes. The proposed system allows to utilize the classical radio frequency (RF) radar target tracking tools  in nanoscale regime for optical signals to track nanoscale and microscale biological units. It provides a basic tool and capability for nanoscale tagging, identification and \textit{ multiple particle tracking (MPT)} applications.

\begin{figure*}[!ht]
\centering
\includegraphics[width=16.0cm]{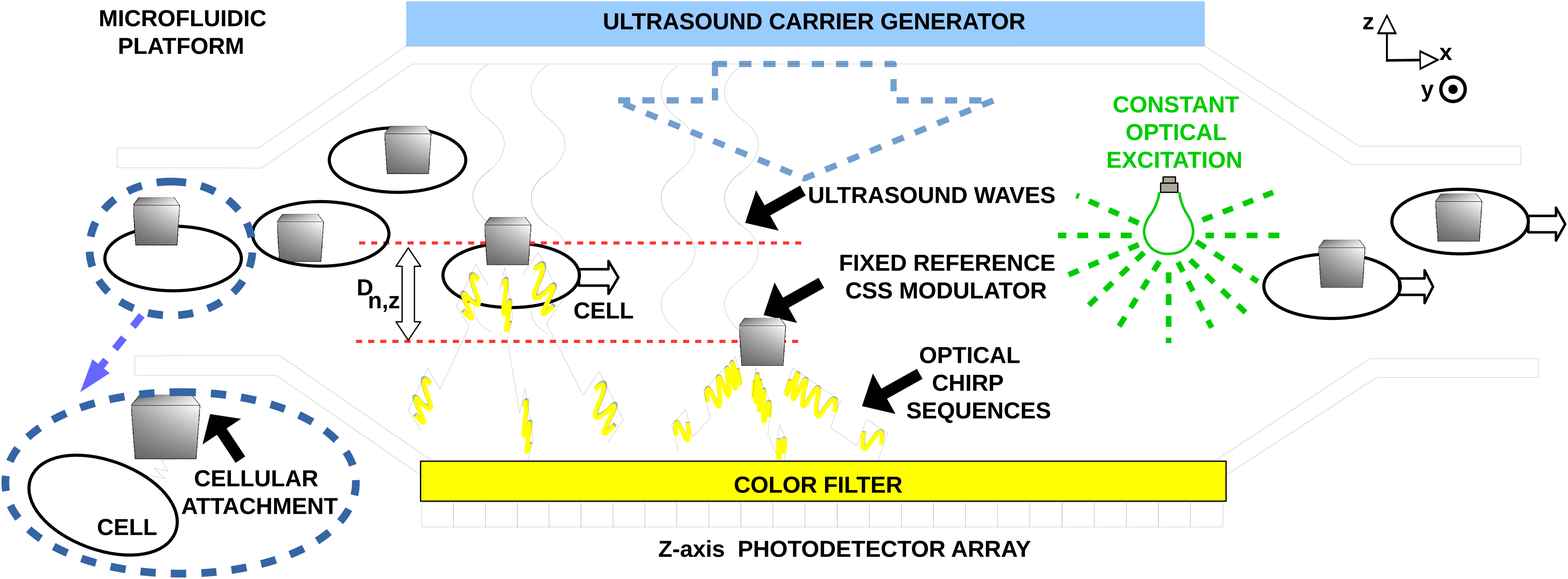}\\
\caption{  \textit{CSSTag}: a nano-biological multiple particle tracking system by utilizing TDOA and CSS based tracking and identification.}
\label{figtag3}
\end{figure*} 

State-of-the-art optical imaging based SPT systems suffer from problems due to resolution, detection speed and multiple cell discrimination in a highly complex cellular topology preventing to get accurate imaging \cite{spt1}. On the other hand, the collection of high resolution data and their fast analysis are highly costly for large numbers of particles requiring artificial intelligence based pattern recognition systems. In this article, the  imaging capabilities  are improved with a signaling based architecture by providing simplicity, the capabilities of classical radar algorithms, the biocompatibility of graphene and  low complexity MPT architecture. 
 
Optical tracking systems promise label free diagnostics such as single particle interferometric reflectance imaging sensor in \cite{interf}, optofluidic intracavity spectroscopy technique spectrally discriminating single biological cells in \cite{ofis} and 2-D optical scanner for imaging up-converting phosphor labels in immunoassays in \cite{osc}. In \cite{timestretching}, optical time-stretch ultrafast imaging is proposed to read barcodes within nanoseconds while requiring spectrally structured active illumination. CSSTag can be utilized in full compatibility  with the high speed fluorescence microscopy and  tracking systems by generating optical tags to be detected by these systems. In addition, point-like excitation of fluorescing molecules requires focusing laser beams  as a challenge for large number of particles \cite{sfim}.   The problems related to the laser sources, complicated set-ups and low imaging penetration depth are solved with LED sources positioned freely in the microfluidic environment. Besides that available image processing methods can be used with CSSTag to improve their performances. For example, in \cite{im2}, significant challlenges are emphasized for low signal-to-noise ratio (SNR) spot detection in fluorescence microscopy which can be experimented with CSSTag. Heavy noise and background intensity changes are handled with random finite set Bayesian filtering for time-lapse cell microscopy  in \cite{im3}. An automatic tracker with a Bayesian probabilistic framework is proposed in \cite{im4}. CSSTag improves the proposed methods by providing  time-varying spread spectrum signaling capability. Manufacturing and experimenting CSSTag system are left as  future works. The theoretical design and simulation performance presented in this article support  future efforts to realize experimental prototype.

In this article, an acousto-optic nanoscale radar system design shown in Fig. \ref{figtag3} is proposed by utilizing the recently introduced  VFRET mechanism in \cite{bg1} to generate CSS sequences with resistance to Doppler effects and higher interference rejection capability \cite{lora}. The modulator combines the advantages of graphene membranes with ultra-low weight, high Young's modulus, strength and wideband resonance frequencies and QDs with broad absorption spectrum, large absorption cross-sections, tunable emission spectra, size dependent emission wavelength, high photochemical stability and improved quantum yield. The hybrid structure converts the acoustic excitation to the periodic optical signals detected by photodetector arrays without requiring high resolution camera or high illumination power. SPT is achieved with time difference of arrival (TDOA) based hybrid system while \textit{TF modulation based radar signaling}  mechanism is brought into the nanoscale regime for MPT. 

In this work, the novel contributions, achieved for the first time, are listed as follows: 
\begin{itemize}
\item A nanoscale optical communications modulator converting acoustic excitations to optical emissions utilizing CdSe/ZnS QDs and vibrating graphene membranes, and a novel mechanical design for acousto-optic frequency multiplication and optical CSS sequence generation. 
\item  A nanoscale optical radar with low computational complexity and  high speed tracking capability with the large numbers of particles and operation with low level of light sources without spatial alignment.  
\item A tagging system improving the state-of-the-art   fluorescence imaging based tracking systems with optical signaling tags in a compatible design to be utilized in these systems with the extended capabilities for diagnostic microfluidic and in-body applications such as cell identification, cytometry and cellular imaging. 
\item Numerical and Monte-Carlo simulations for two particle tracking experiment with micrometer size modulators having weights of several picograms and low level of excitation sources compared with laser sources showing high performance with low SNR values between $-7$ dB and $10$ dB. 
\end{itemize}

The remainder of the paper is organized as follows. In Section \ref{s2}, VFRET system model is summarized while  the device structure and fabrication are described in Section \ref{s3}. In Section \ref{s4}, CSS sequence generation is modeled and CSSTag MPT mechanism is introduced. Then, in Section \ref{s5}, the challenges are discussed. The numerical and Monte-Carlo simulations are presented in Section \ref{s6} while the conclusions are given in Section \ref{conclusion}.

\section{Vibrating FRET Modeling}
\label{s2}

 \begin{figure*}[!t]
\centering
\includegraphics[width=18cm]{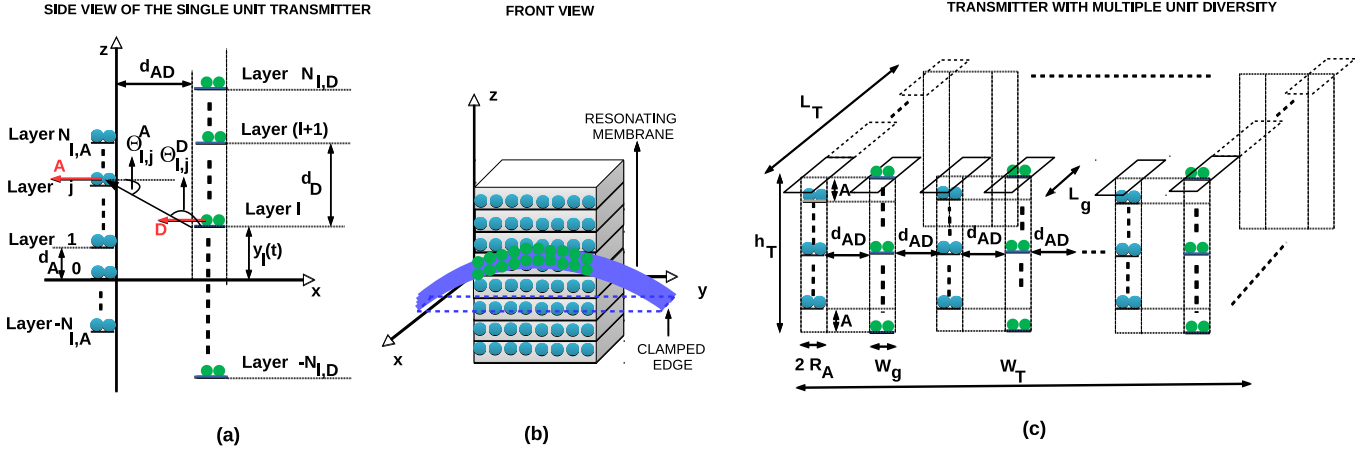}\\
\caption{ Single unit of the frequency multiplier structure with (a) side view,  (b) front view and (c) the multi-unit diversity in a stacked architecture which are extended and modified versions of the simple units defined in \cite{bg1, bga, bgb, bgc}.}
\label{figMult}
\end{figure*}

Doubly clamped single layer graphene (SLG) beams are modeled with the resonance frequency of $f_{r} = (1/ \, 2 \, L_g) \sqrt{F_T \, / \, (\rho_g  \, W_g)}$ under a specific strain level $S_m \, (\%)$ due to applied periodic excitation sound pressure where  $F_T$ is the pre-tension in (N/m) and,  $L_g$, $W_g$ and  $\rho_g$ are the length, width and the mass density of the resonator, respectively \cite{bg1, mn1}. The resonance frequency in a fluid medium compared with the models defined for the  vacuum is assumed to be approximated by $f_{w} = f_r \, / \, \sqrt{1 \, + \, \Gamma_w \, \rho_w \, L_g \, / \, (\rho_g \, h_g)}$ where $\Gamma_w$ is a constant factor, $\rho_w$ is the mass density of the fluid medium and $h_g$ is the graphene membrane thickness \cite{kwak}. Furthermore, the resulting resonance frequency $f_0$ due to the QD load on the graphene drum  is approximated  by  $f_0 = f_w  \, \big(1 \, -   \,\vert \partial m \vert/ \, (2 \,m_d ) \big)$ where $m_d$ is the mass of the graphene drum and $\vert \partial m \vert$ is the load on the drum.  $T_p \equiv 1/f_0$  denotes the time period of the resonator \cite{bg1}. Then, the number of photons emitted in VFRET mechanism introduced in \cite{bg1} for low power excitations and the rates smaller than lifetime limited maximum is modeled as follows:  
\begin{equation}
N_{FRET}= \frac{I{_{D}}\,D_{B}\,\sigma _{D} \, \Phi{_{A}} \,E_{FRET} \, N_D \,\Delta t}{h_p \, f_D^a}  
\end{equation}  
where $I{_{D}}$ is the light intensity in ($W/m^2/nm$) for donor excitation, $D_B$ is the bandwidth of donor excitation in $(nm)$, $\sigma_{D} = D_{ext}\, 3.825 \,10^{-25}$ is the absorption cross section of donor excitation in ($\mbox{m}^2$), $D_{ext}$ is the donor extinction coefficient in $(\mbox{M}^{-1} \,\mbox{cm}^{-1})$, $\Phi{_{A}}$ is the quantum yield of acceptors, $N_D$ is the number of donor molecules, $\Delta t$ is the photon counting time set to larger than $t_{D} \, + t_{FRET} \, + t_{A}$ which is the total time on the order of nanoseconds for photon emissions from the acceptor including the donor excitation time $t_{D}$, $FRET$ time $t_{FRET}$ and the acceptor emission life time $t_{A}$,    $f_D^a =  c \, / \, \lambda_D^a$ is the donor excitation frequency,  $\lambda_D^a$ is the donor excitation wavelength,  $h_p = 6.62 \, 10^{-34}$ $(\mbox{J} \times \mbox{s})$ is Planck's constant and $c = 3 \, 10^{8}$  m/s is the velocity of light. $t_{FRET}$ is much smaller compared with $t_{A}$ and $\Delta t$ is set to $2 \times t_A$  to obtain an averaging behavior for the exponential decay profile of the emission,  blinking, flickering and Poisson statistics of photon arrivals  \cite{qd1}. FRET efficiency $E_{FRET}$ is defined as $E_{FRET} \, = \,  k_{A} \, R_{0}^{6} \, / \, (d_{FRET}^{6} \,+ \,k_{A} \, R_{0}^{6})$ where  $k_{A}$ is the number of acceptors per donor, $d_{FRET}$ and $R_{0}$ are the donor-acceptor and F{\"{o}}rster distances, respectively \cite{bg1}. The acoustic excitations due to sound pressure levels on graphene membrane are converted to optical emissions due to FRET mechanism between the donor cluster attached on graphene and the acceptor at a fixed position. The generated light intensity has the same period of the acoustic resonance while the intensity of excitation is tunable with the number of donors and acceptors, the resonance frequency, thickness of the graphene layer ($h_g$) and the strain level ($S_m$) by the sound pressure as discussed in detail in \cite{bg1}. In this article, SLG is utilized to create a multi-layer modulator structure with ultra-light weight properties. Next, the model defined in \cite{bg1} is significantly improved to realize optical frequency modulation (FM), frequency multiplier and CSS sequence generation.

\section{Acousto-optic FM Modulator and  Multiplier}
\label{s3}

The novel acousto-optic modulator structure with acousto-optic frequency multiplication capability is shown in Fig. \ref{figMult}. Next, the properties of the device are explored by specifying the effects of dynamic F{\"{o}}rster radius, the diversity combining architecture and VFRET multiplier emission modeling.

\subsection{Dynamic F{\"{o}}rster Radius}
\label{s3s1}
 
The dynamic F{\"{o}}rster radius modeled with $R_{0,l,j}^6(t) = R_0^6 \, \kappa^2_{l,j}(t)$ due to periodic oscillation of the donor molecules depends on the relative orientation of the donor and the acceptor (shown in red) as the donor resonates along the acceptor building  as shown in Fig. \ref{figMult} where $\kappa^2_{l,j}(t) = \big(\cos \theta_{l,j}^{DA}(t)   - 3 \, \cos \theta_{l, j}^D(t)  \, \cos \theta_{l,j}^A(t)    \big)^2$ for the donor-acceptor (D-A) pairs on the $l$th and $j$th layers, respectively, $R_0$ is the reference F{\"{o}}rster distance measurement for D-A pairs at the same positions with respect to y and z axes \cite{dfr}. The angle $\theta_{l,j}^{DA}(t)$ is assumed to be zero for the  D-A pairs at the same position in the y-axis. $d_A$ and $d_D$ denote  the inter-layer distances between the acceptor and the donor layers, respectively, where $d_{AD}$ is the fixed distance in the x-axis between the groups of the acceptor and the donor layers.  The angle $\theta_{l, j}^D(t)  = \pi- \theta_{l, j}^A(t)$ is calculated as $\tan^{-1}(\vert j \, d_A - y_l(t) \vert \, / \, d_{AD})$ for the current position $y_l(t)$ of the resonating $l$th graphene layer. $\kappa^2_j(t)$ is changing between one and four from the long distance to the  direct coupling case in the close neighborhood. Next, the stacked device structure with a transmitter diversity is presented. 
 
\subsection{Transmitter Diversity with the Stacked Resonators}
\label{s3s2}

The single unit device shown in Fig. \ref{figMult}(a) has total number of $2 \times N_{l,A} +1$ and $2 \times N_{l,D} + 1$ layers of acceptors and donors, respectively, and the donor and acceptor layers have the layer indices in the intervals $\left[ -N_{l,A}, N_{l,A} \right]$ and $\left[ -N_{l,D}, N_{l,D} \right]$, respectively. It is assumed that the vertical positions of the  donor and acceptor layers  with the index zero are the same in the z-axis at stable conditions without any resonance.  The diversity combining stacked device volume is assumed to be $h_T \times W_T \times L_T$ where $h_T$, $W_T$ and $L_T$ are the height, width and the length of the multiplier device  composed of the stacked unit resonator blocks as shown in Fig. \ref{figMult}(c).  $N_{l, A}$ is determined with respect to the maximum strain  in order to create a homogeneous light emission among different acceptor layers since the layers at the highest and lowest vertical positions experience different  coupling with the boundary layer donors.  The homogeneity is satisfied  by allowing the donor layers at the highest and lowest vertical positions to have a separation distance to the closest acceptor layer equal to the maximum amplitude of the vertical resonance amplitude, i.e., $A$, as follows:
\begin{equation}
N_{l, A} = (h_T -2 \, A) \, / \, (2 \, d_A) 
\end{equation}
where $d_A \equiv k_{\Delta  A} R_0$ depends on the predefined constant  $k_{\Delta  A}$ and larger than $R_{A} + h_{f}$, $R_A$ is the acceptor QD diameter  and $h_f \ll d_A$ is the thickness of the rigid and non-resonating structure that the acceptors reside in. $N_{l,D}$ is set to $h_T \,  / \, (  2\, d_D)$ where $d_D  \equiv  k_{\Delta  D} R_0$ is chosen larger than $h_g \, + \,R_{D} + d_{\lambda}$ by tuning the variable $k_{\Delta  D}$ where $R_{D}$ is the donor QD diameter.  The extra factor $d_{\lambda}$ is the maximum vertical displacement due to the phase difference of the excitation between neighbor layers, i.e., $d_{\lambda}  \equiv \underset{t}{\mbox{max}} \lbrace A \, \sin( \omega_0 \, t + 2\, \pi \,d_D \, / \, \lambda_e) -  A \, \sin( \omega_0 \, t) \rbrace$  where $\lambda_e = v_e \, / \, f_0$ and $v_e$ are the wavelength and the velocity of the excitation signal in the medium, e.g., sound velocity in liquid medium. The  single unit device has the volume of $h_T \times (W_g + 2 \,d_{AD} + 2 \, R_{A}) \times L_g$ assuming that $W_g > 2\, R_{D}$. Then, the total number of the unit devices is found approximately by $\big(W_T \,/ \, (W_g + 2 \, d_{AD} + 2 \, R_{A}) \big) \times (L_T \, / \, L_g)$ where $R_A > R_D$ is assumed due to the utilized set of CdSe/ZnS QDs in this article. Next, the frequency multiplier effect is theoretically modeled.
 
\subsection{Frequency Multiplication Modeling}
\label{s3s3} 

Assuming that the current position of the donors on the $l$th layer is given by $y_l(t) = l\,d_D + A \, \sin(\omega_0 \, t \, + \, 2\, \pi \,  l\,d_D \, / \, \lambda_e )$ in a single unit device, the emitted light intensity by the acceptors on the $j$th layer at the time interval between $i \, \Delta t$ and $(i+1)\, \Delta t$ is trivially calculated as approximately equal to the following:
\begin{eqnarray}
I_{l,j}(i \, \Delta t)  =  I_{D} \, D_{B} \, \sigma _{D} \, \Phi{_{A}} \, \Delta t \, N_D   \hspace{0.5in} && \nonumber  \\ 
 \times \, \frac{k_{A} \, R_0^6 \,  }{\frac{(d_{AD}^2 + \vert y_{l,j}(i \, \Delta t) \vert^2)^4 }{ 4 \, d_{AD}^2 + \vert y_{l,j}(i \, \Delta t) \vert^2 }\,+ \,k_{A} \, R_0^6 \,  } && 
\end{eqnarray}
where $y_{l,j}(i \, \Delta t) = j \, d_A -y_l(i \, \Delta t)$. $N_D$ is given by two single chains of QDs placed on the two sidelines of the graphene layer which are not clamped as shown in Fig. \ref{figMult}(b) as follows:
\begin{equation}
N_D = 2 \times \frac{L_g}{\alpha_{QD} \, D_{QD}} 
\end{equation}
where the QDs  are placed on a short line compared with $L_g$, i.e., $L_g \, / \, \alpha_{QD}$ with $\alpha_{QD} \gg 1$, such that the donors on the same layers experience approximately the same displacement as the resonator vibrates. 

It is assumed that $d_{D}, d_{A} > R_0$ such that at a single time interval $\Delta t$, each acceptor is excited by only a single donor with the closest distance at the z-axis. Then, the total number of emitted photons in the time interval $i \, \Delta t$  is given  as follows:
 \begin{align}
\begin{split}
n_{p, A}(i \, \Delta t)=  &\sum_{j = -N_{l,A}}^{N_{l,A}} \underset{l \in [-N_{l,D}, N_{l,D}]}{\mbox{max}}  \bigg \lbrace \frac{I_{l,j}(i \, \Delta t)}{h_p \, f_D^a}  \bigg \rbrace   \\
&\times \, \frac{W_T}{W_g + d_{AD} + 2 \, R_{A}} \,  \frac{L_T}{L_g}  \,    
\end{split}
\end{align}

The acousto-optic frequency multiplication occurs with the multiple periodic emissions in each acceptor layer due to the large vibration amplitude of $A \gg d_A$  such that the donor in a single layer visits multiple acceptor layers in  a single acoustic vibration period $T_p$. As a result, low frequency acoustic signals are converted to high frequency optical signals with a multiplication factor limited by the total photon emission time $\Delta t$, the resonance amplitude and the distance between the layers. Assume that the position of the membrane on the $l = 0$ layer for a sinusoidal excitation is approximately given by $A \, \omega_0 \, \sin(\omega_0 \, t)$ where the membrane is at the same level with the acceptor layer on the $j = 0$ layer at $t =0$. Then,  the period of the emitted photon intensity between the first two acceptor layers as the donor membrane passes is approximately  given by $T_{min}    \equiv t_{min,1} - t_{min,0}$ where $t_{min,1}  =  (1 \, / \, \omega_0) \sin^{-1}( (d_{A}+ R_0) \, / \, A )$ and $t_{min,0}   =  (1 \, / \, \omega_0) \sin^{-1}(  - R_0 \, / \, A )$ resulting in the maximum frequency emission of $f_{max} =  f_0 \times M_{max}$ where   the multiplication factor is given as follows:
\begin{equation}
M_{max}  \equiv  \frac{ 2 \, \pi}{\sin^{-1}(\frac{d_{A}+ R_0}{ A}  ) \, + \, \sin^{-1}( \frac{R_0}{ A})}
\end{equation}
Similarly, the minimum frequency photon emission period occurring at the peak amplitude of the resonating membrane where the velocity drops to zero is approximated by  $T_{max} \approx t_{max,1} - t_{max,0}$ where $t_{max,1}  = (1 \, / \, \omega_0) \pi \, / \, 2$ and   $t_{max,0}   =   (1 \, / \, \omega_0) \sin^{-1}( 1  - (d_{A}+ R_0) \, / \, A )$. Then, the minimum multiplication factor drops to the following:
\begin{equation}
M_{min}  \equiv  \frac{ 2 \, \pi}{\pi \, / 2 -  \sin^{-1}( 1  - (d_{A}+ R_0) \, / \, A )}
\end{equation}
A CSS sequence  with a nonlinear FM modulation between $f_{min} = M_{min} \times f_0$ and $f_{max} = M_{max} \times f_0$ is generated compared with hyperbolic and sinusoidal CSS sequences in literature \cite{css1}. Next, the fabrication of the proposed device is discussed.

\subsection{Device Fabrication}
\label{s3s4}
The proposed architecture should be experimentally verified based on the presented unique mechanical design. The recent works on graphene are promising to manufacture complex designs with nanometer accuracy.  The construction of multiple arrays of resonating graphene membranes is experimentally feasible based on the studies in \cite{la},  dropcasting of QDs on graphene membrane is achieved in \cite{et, qq1} and  the frame structure where the acceptors reside and the graphene membranes are clamped can be chosen as carbon membranes with nanoscale thicknesses allowing complex layered designs \cite{cmemb}. Besides that, the attachment on cells or biological units is feasible based on the experimental studies showing the cells adhering to graphene oxide (GO) membrane steadily without harmful toxicity \cite{ bg1, blink1, bg2}.  Next, CSS modulator structure to generate different sequences and the utilization for particle tracking tasks are discussed.

\section{Chirp Spread Spectrum Modulator and Nanoscale Particle Tracking}
\label{s4}

\begin{figure}[!t]
\centering
\includegraphics[width=3.5in]{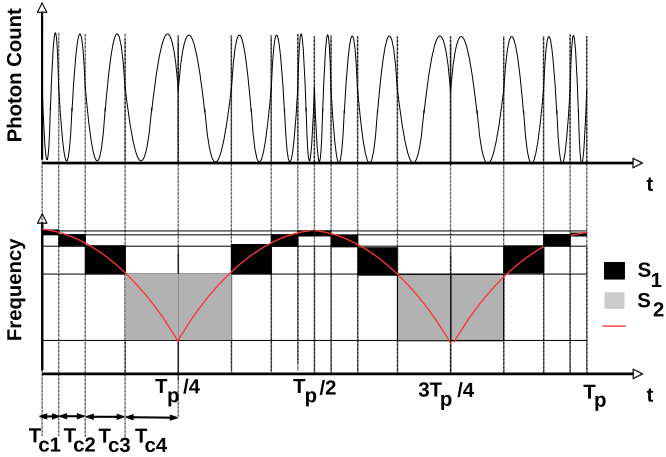}\\
\caption{ A sample acceptor emission photon count waveform with chirped signal structure and TF spreading based division of the waveform into distinct CSS sequences $S_1$ and $S_2$ which can be realized with a special geometrical design of donor and acceptor placements.}
\label{figtag2}
\end{figure}

Orthogonal CSS sequences are generated for multiple particle tracking networks with unique demodulation capability at the optical receiver. Assume that, a set of chip sequences is generated by uniquely positioning the acceptor layers with a number of layers $M$ and sequence index $k \in [1, K]$ denoted by $Cp_{k} = \lbrace j_{k, 1}, \, j_{k, 2}, \hdots, j_{k, M}; j_{k,n} \in  [ -N_{l,A}, \, N_{l,A}] , \, n \in [1, M]\rbrace$ resulting in the following emission chirp sequence:
 \begin{align}
\begin{split}
S_{k}(i \, \Delta t)=& \sum_{j  \in Cp_{k}} \underset{l \in [-N_{l,D}, N_{l,D}]}{\mbox{max}}  \bigg \lbrace \frac{I_{l,j}(i \, \Delta t)}{h_p \, f_D^a}  \bigg \rbrace     \\
& \times \, \frac{W_T}{W_g + d_{AD} + 2 \, R_{A}} \, \times  \, \frac{L_T}{L_g}  
\end{split}
 \end{align}
Two example sets of CSS sequences having both time and frequency spread are shown in Fig. \ref{figtag2} where two sequences are orthogonal in both time and frequency domain providing a significantly powerful chip generation mechanism. It is observed that the duration and the frequency of chip sequences denoted with $T_{ck}$ for $k$th  time interval are different due to the different velocity of the donor while passing through equivalently separated acceptor layers. The frequency of the chips follows the instantaneous frequency of the resonating donor membrane. Next, the generated CSS sequences are utilized firstly for SPT and then for MPT purposes.

\subsection{Single Particle Tracking}
\label{s4s1}

The conventional methods utilized for target tracking, e.g., in a transparent microfluidic cell tracking platform, are applied for the proposed nanoscale system such as hybrid acousto-optical TDOA with acoustic and optical excitations to find the positions of the nano-biological objects that the modulator is attached as shown in Fig. \ref{figtag3}. The position is found by checking the TDOA with a reference object as follows:
\begin{equation}
D_{k, z} = (\tau_{k,z} - \tau_{0,z})\, / \,  v_e
\end{equation}
where z-axis coordinate of the $k$th cell or the biological particle is determined, $\tau_{k,z}$ and  $\tau_{0,z}$ are the delays of the detected CSS sequences and the reference modulator at a fixed position, respectively, and $v_e$ is the sound velocity in liquid which can be measured with sensors. The accuracy is determined by the estimation of the velocity of sound in the medium, detection of the first period of the optical CSS sequence and pilot training with a large set of reference modulators positioned in the microfluidic environment. Therefore, the proposed system provides a simple and accurate SPT  borrowing from classical target tracking approaches. The extension to MPT scenario is achieved by generating CSS sequences with different TF supports.

\subsection{Multiple Particle Tracking}
\label{s4s2}

\begin{figure*}[!t]
\centering
\includegraphics[width=17.5cm]{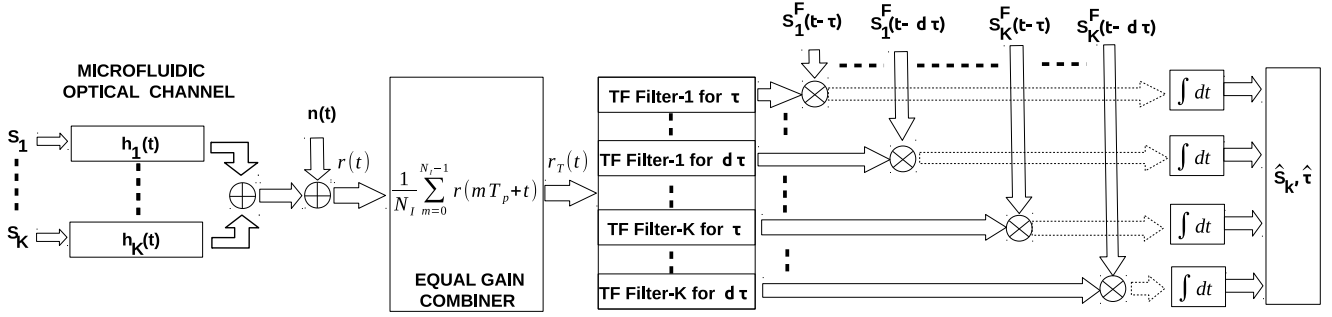} 
\caption{Time-frequency filtering based optimum  multiple particle CSS receiver.}
\label{figtag8}
\end{figure*} 

The tagging denoted by \textit{CSSTag} is achieved by demodulating the sequences of different modulators in a MPT scenario. In MPT, either the particles are attached with the resonators having different ultrasound resonance frequencies or the same frequency $f_0$. In the former case, multiple cells are tracked by exciting them with different frequencies at a single instant and checking the TDOA in the demodulator. However, in the latter case an interfering set of CSS sequences should be demodulated. Then, with the similar approach in the x and y axes, a scanning low-cost system is realized to quickly track low velocity targets. It is assumed that that the cells do not experience high mobility due to the resonating effect of the sound waves, e.g., $\lambda_e = 2.2$ cm for $f_0 \approx 71$ KHz. On the other hand, $f_0$ can be increased with lower $L_g$ to obtain wavelengths comparable with cellular sizes. The proposed system does not utilize sound modulation or ultrasound receiver, therefore, it is a low-cost system design with high accuracy target detection capability for nanoscale communication networks. 

The auto-correlation and cross-correlation of chirp sequences are analyzed to define their robustness with respect to inter-symbol interference (ISI) and narrow band noise. The correlation is defined for two CSS tags with indices $i$ and $j$ having the photon count periodic vectors $\mathbf{S_{i}}$ and $\mathbf{S_{j}}$ as follows:
 \begin{align}
\begin{split}
r_c(\mathbf{S_{i}},\mathbf{S_{j}}, \tau) = &  \frac{1}{N_I }  \, \sum_{m=0}^{N_I-1} \sum_{n=0}^{N_p-1} \frac{S_{i}\big(\upsilon(m,n)\big)}{E_i}   \hspace{0.2in}  \\ 
   & \times \, \frac{S_{j}\big(\upsilon(m,n) \, + \, \tau \big) }{ E_j} 
  \end{split}
\end{align}
where  $\upsilon(m,n) = m T_p \, + \, n\, \Delta t$, $N_I$ different periods are utilized with an equal gain combining (EGC) diversity approach and an estimation method similar to classical periodogram \cite{spectral} to find the average correlation, $E_k$ is the average photon count energy defined as $E_k \equiv  \sum_{n=0}^{N_p-1} S_{k}^2(n\, \Delta t)$ for $k \in [1, K]$, $N_p \equiv T_p/\Delta t$ and cyclic shift is utilized due to periodically oscillating CSS sequences. In the following, the received signal is firstly diversity combined with EGC by utilizing the samples in multiple periods and, then the  waveform denoted by $r_T(t)$ is passed through the correlator or matched filter to estimate the delays.  

Assuming that the channel is not frequency selective for the low bandwidth of the CSS signal and the velocities of the cells are small compared with the speed of sound, a TF receiver architecture is utilized to track the positions of multiple cells  as shown in Fig. \ref{figtag8} including a delay estimation block similar to a RAKE receiver for multi-path multi-user receivers \cite{gold2}. The received optical emission signals  $\mathbf{S_{k}}$ for $k \, \in [1, K]$ pass  through the microfluidic optical channel including the effects of scattering, attenuation and optical absorption \cite{ch1}. However, it is assumed that the channel is not frequency selective, the CSS sequence has a low bandwidth and a large number of $N_I$ periods are utilized to estimate the total emission in a single period of $T_p$ to simply define SNR performance metrics based on the summation of individual photon counts in the simulation studies. The proposed framework can be easily extended to time-varying and frequency selective optical channels. The received signals are passed through EQC to obtain $r_{T}(t)$ denoting the received signal waveform in a single period. Then, the corresponding TF CSS filter is applied on the received signal for each discrete time delay in $[0, d \times \tau]$ and for each CSS sequence where $\tau$ and $d$ are determined with respect to the desired resolution and the maximum distance to the reference modulator in the microfluidic chip. The results are correlated with TF filtered and delayed sequences denoted by $\mathbf{S_{k}^F}(t - j \, \tau)$ for $j \in [1,d]$ and $k \in [1, \,K]$. Then, the estimate $\widehat{\tau}_k$ is obtained for each index $k$. The computational complexity is discussed next.

\subsection{Computational Complexity of MPT}
\label{s4s3}

The computational complexity  includes $K\,  \times \, d \, \times \, N_p$ multiplications in time filtering, $ K \, \times \, d  \, \times \, N_p  \, \mbox{log}N_p$ operations in frequency domain filtering and,  $ K\,  \times\,  d \, \times  \,N_p$ multiplications and $K \times d$ additions for the correlation blocks. It is possible to realize the set of correlations with complexity on the order of $O(K\, \times \, N_p\, \times\, \mbox{log} N_p)$ instead of $O(K \, \times d \, \times \, N_p)$ by utilizing Fast Fourier Transform (FFT) based approaches where $d \, \leq  \,N_p$. The application of  more complex filtering and  detection algorithms in classical target tracking literature are left as future works to optimize the performance. Next, the simplifications in the channel model for simulation purposes and the definition of SNR are presented.

\subsection{Channel and SNR Modeling}
\label{s4s4}

Assuming a frequency flat and non time-varying optical channel response $H(f_0,t)$ for the duration of $N_I \times \Delta t$ on the order of several microseconds, the  received signal $r(t)$ is represented as follows:
\begin{equation}
r(t) = \sum_{k = 1}^{K} H_k(f_0,t) \,  S_{k}(t - \tau_k) + n(t)
\end{equation}
where the channel response of each channel is denoted by $h_k(t)$ with frequency flat response of $H_k(f_0, t)$ under additive white Gaussian noise $n(t)$ assumption. In Section \ref{s6}, the performance is simulated for varying total SNR defined as follows:
\begin{equation}
SNR   \equiv \frac{1}{K^2} \bigg \vert \sum_{k = 1}^{K}  \mu_k    \bigg \vert^2 \frac{ \vert H(f_0,t) \vert^2}{ \sigma_{n}^2} 
 \end{equation} 
where a simplification is utilized for equal channel responses $H(f_0, t)$ to provide a measurement metric for SNR and the mean number of photons for CSS $Tag_j$ is denoted by $\mu_k \equiv (1 \, / \, N_p)\sum_{m =0}^{N_p - 1} S_{k}(m \, \Delta t)$. The Gaussian noise is assumed to have equal mean $(\mu)$ and variance $(\delta_n^2)$ denoted by $N(\mu,\delta_n^2)$ as discussed in \cite{poisson} by approximating the Poisson process of photon arrivals  at each time interval $\Delta t$  resulting in a more accurate and worst case SNR definition compared with $E_k^2 \, / \, \delta_n^2 < \mu_k^2 \, / \, \delta_n^2$. The performance obtained with a specific level SNR defined with respect to $\mu_k^2 \, / \, \delta_n^2$ is converted to a lower SNR defined with respect to  $E_k^2 \, / \, \delta_n^2$.
 
It is left as a future work to extend the framework to include Doppler velocity estimation, to optimize for more efficient calculations and to combine with the classical radar and target tracking solutions  utilizing Fractional Fourier Transform based TF filtering with slicing on TF plane \cite{haldun}. More complex CSS sequences allow a code division multiple access (CDMA) architecture for the MPT radar. The proposed simultaneous  TF filtering system is robust to noise or ISI, and better design of CSS sequences and the performance studies are open issues.

\section{Challenges and Open Issues}
\label{s5}

There are a set of challenges for the realization of the proposed particle tracking system and modulators, and a set of open issues promising to significantly  improve the simulated theoretical performances in this article as follows. The simplicity of the proposed theoretic model and device, and the simulation performance support further efforts to realize CSSTag prototype and perform experiments. Experimenting and verifying both VFRET phenomenon and CSSTag are left as future works.

\subsection{QD Related Issues}
\label{s5s1}

Blinking and flickering problems observed in QDs as discussed in \cite{blink1, bio2} are averaged due to the significant number of D-A pairs in a transmitter utilizing diversity of multiple resonators and layers, and utilization of high number of repetitive periods with EGC based estimation. Photo-bleaching  is not a significant concern due to the high durability of QDs compared with other organic materials and low photon count for each QD under low level of light intensity \cite{blink1}. On the other hand, the exponential decay profile of acceptor fluorescence emission is approximated by counting the photons in an interval of $2 \, t_{A}$. It will be further analyzed as a future work to find the speed limits of frequency multiplication for high speed nanoscale optical applications in addition to the in-body and microfluidic applications, e.g., LED based CSS high speed modulators for visible light communications (VLC). The modulation speed is limited by the fluorescence lifetime which can be improved with different materials, e.g., fast radiative lifetimes and FRET of CdSe nanoplatelets (NPLs) on the orders of tens of picoseconds \cite{npl}, or with recent studies showing the decrease in lifetime for CdSe/ZnS QDs dropcasted on graphene and $\mbox{MoS}_2$ layers \cite{et, qq1}.  

\subsection{Optical Channel Modelling and Receiver Performance}
\label{s5s2}

The efficiency of the photo detection can be improved with PMTs providing $1000 \times$ gains to further improve the sensitivity of the proposed device. On the other hand, the effects of the channel and the collection efficiencies of the photodetectors should be modeled with more accuracy where a frequency flat model is assumed in this article with a large number of tag periods $N_I$ and small bandwidth in the CSS sequences. Modeling the attenuation, absorption and scattering of the microfluidic channel is necessary to exactly model the SNR performance at the receiver. In this article, both high and low SNR regimes are considered to simulate the significant effects of the channel. If there is frequency-selective response due to special bio-organic topology between the photodetector and the cell, then  the response can be utilized to tailor specific TF tags with special arrangements of D-A pairs.

\subsection{Toxicity and Biocompatibility}
\label{s5s3}

Toxicity of CdSe/ZnS QDs is discussed in \cite{blink1}. It is tolerated due to low interaction time of the cells with the modulators and the potential for special graphene or carbon membrane enclosure preventing the QDs to disperse out of the device. Besides that, graphene is a biocompatible material with significantly increasing number of experimental studies creating methods for making it more suitable for biomedical applications \cite{bg2}.
 
\section{NUMERICAL SIMULATIONS}
\label{s6}

\setlength\tabcolsep{2 pt}  
\newcolumntype{M}[1]{>{\centering\arraybackslash}m{#1}}
\renewcommand{\arraystretch}{1.3}
\begin{table}[t!]
\caption{Simulation Parameters}
\begin{center}
\scriptsize
\begin{tabular}{ |m{2.1cm}|m{1.4cm} |m{2.1cm}|m{1.4cm}| }
\hline
 \multicolumn{1}{ |c| }{PARAMETER} & \multicolumn{1}{ c| }{VALUES} &  \multicolumn{1}{ c| }{PARAMETER} & \multicolumn{1}{ c| }{VALUES} \\ \hline
 \multicolumn{1}{ |c| }{$\rho_g$} & \multicolumn{1}{ c| }{ $ 2200$  ($\mbox{kg}/\mbox{m}^3$)} & \multicolumn{1}{ c| }{$F_T$} & \multicolumn{1}{ c| }{ $1$ N/m  } \\ \hline
\multicolumn{1}{ |c| }{  $ K (\# \mbox{Cells})$} & \multicolumn{1}{ c| }{  2 }  &   \multicolumn{1}{ c| }{$k_{A}$} & \multicolumn{1}{ c| }{ $1$ } \\ \hline
  \multicolumn{1}{ |c| }{$m_{A}$ } & \multicolumn{1}{ c| }{ $3.04 \, 10^{-19}$ (gr)  } &  \multicolumn{1}{ c| }{  $m_{D}$} & \multicolumn{1}{ c| }{ $1.46 \, 10^{-19}$ (gr)}\\ \hline 
    \multicolumn{1}{ |c| }{$\lambda_D^a$ } & \multicolumn{1}{ c| }{ $540$ (nm)  } &  \multicolumn{1}{ c| }{ $\lambda_A^e$ } & \multicolumn{1}{ c| }{ $580$ (nm)}\\ \hline
  \multicolumn{1}{ |c| }{$D_{ext}$} & \multicolumn{1}{ c| }{$0.02 \, (\mbox{M}^{-1}  \mbox{cm}^{-1})$ }  & \multicolumn{1}{ c| }{$I_{D}$} & \multicolumn{1}{ c| }{100, 1000 ($W/m^2/nm$)} \\ \hline
   \multicolumn{1}{ |c| }{$R_{D}$} & \multicolumn{1}{ c| }{  $2.9$ (nm)} & \multicolumn{1}{ c| }{$R_{A}$} & \multicolumn{1}{ c| }{ $3.7$ (nm) } \\ \hline
 \multicolumn{1}{ |c| }{$D_{B}$} & \multicolumn{1}{ c| }{  $40$ (nm)} & \multicolumn{1}{ c| }{$R_0$} & \multicolumn{1}{ c| }{ $6.63$ (nm) } \\ \hline
 \multicolumn{1}{ |c| }{$\Phi_{A}$ } & \multicolumn{1}{ c| }{ $0.6$  }  & \multicolumn{1}{ c| }{$d_{AD}$ } & \multicolumn{1}{ c| }{  $2$ (nm)}\\ \hline
  \multicolumn{1}{ |c| }{$  k_{\Delta  A},   k_{\Delta  D}$ } & \multicolumn{1}{ c| }{ $4$, $88$ }  & \multicolumn{1}{ c| }{$\alpha_{QD}$  } & \multicolumn{1}{ c| }{  $50$ }\\ \hline
   \multicolumn{1}{ |c| }{$L_g$ } & \multicolumn{1}{ c| }{  $10$ $\mu$m }  & \multicolumn{1}{ c| }{$W_g$ } & \multicolumn{1}{ c| }{  $20$ (nm)}\\ \hline
    \multicolumn{1}{ |c| }{$h_T$ } & \multicolumn{1}{ c| }{ $10$  $\mu$m }  & \multicolumn{1}{ c| }{$h_g$} & \multicolumn{1}{ c| }{  $0.335$ (nm) }\\ \hline
     \multicolumn{1}{ |c| }{$W_T = L_T$  } & \multicolumn{1}{ c| }{  $10$ $\mu$m }  & \multicolumn{1}{ c| }{$ S_m (\%)$ } & \multicolumn{1}{ c| }{  $0.005$ -- $1$}\\ \hline 
\multicolumn{1}{ |c| }{ $SNR$ } & \multicolumn{1}{ c| }{ $-7$ to $10$ dB} &  \multicolumn{1}{ c| }{$v_e$ } & \multicolumn{1}{ c| }{$1543$ (m/s) }\\ \hline 
\multicolumn{1}{ |c| }{ $\Gamma_w$ } & \multicolumn{1}{ c| }{ $0.74$  } &  \multicolumn{1}{ c| }{$\rho_w$ } & \multicolumn{1}{ c| }{$1000$ (kg/$\mbox{m}^3$) }\\ \hline 
\end{tabular}
\end{center}
\label{table2}  
\end{table}
\renewcommand{\arraystretch}{1}
\setlength\tabcolsep{6 pt}  
 
The performance of the proposed MPT system is simulated with the  parameters given in Table \ref{table2} for both the frequency multiplier architecture and the target tracking architecture for  $K = 2$  particles or the cells with CSS tags denoted by $Tag_1$ and $Tag_2$. The graphene membrane parameters of $\rho_g$ and $F_T$ are set to the practical parameters defined in literature \cite{bg1}. The donors and the acceptors are chosen as CdSe/ZnS QDs with diameters of $R_{D} = 2.9$ nm and $R_{A} = 3.7$ nm, respectively, with donor absorption and acceptor emission wavelengths of $\lambda_D^a = 540$ nm and $\lambda_A^e = 580$ nm, respectively, and the parameters of $D_{ext}$, $D_B$, $\Phi_{A}$ and $R_0$  are defined in \cite{bg1}. Their weights $m_{D}$ and $m_{A}$ are calculated based on the reference weight of the QD with $2.5$ nm diameter simulated in \cite{bg1} by assuming QDs as homogeneous spherical structures.  The membrane dimensions are set to $W_g = 20$ nm  and $L_g = 10 \, \mu$m for SLG  with the thickness $h_g = 0.335$ nm. The resonance is in the ultrasound regime with $f_0  \approx 71$ KHz in the fluid environment assumed as water with $\Gamma_w \approx 0.74$ given in \cite{kwak}, the mass density of $\rho_w  \approx 1000$ (kg/$\mbox{m}^3$) and $ v_e \approx1543 $ m/s.  The total device dimension parameters are set to $h_T = W_T = L_T = 10 \, \mu$m suitable to be utilized in microscale biological units. $\alpha_{QD} = 50 \gg 1$ results in a resonating cluster of QDs at the similar vertical heights. $k_{\Delta A}$ and $k_{\Delta D}$ are set to four in the multiplier device to create independent D-A pairs while $k_{\Delta D}$ is set to $88$ in the radar tracking system to realize different CSS optical emission sequences. The excitation light intensities are set  to $100$ ($W/m^2/nm$) and $1000$ ($W/m^2/nm$) for the multiplier and  tracking cases, respectively, with more powerful light sources for tracking while $1000$ ($W/m^2/nm$) corresponding to $4$ ($W/cm^2$) in QD bandwidth of $40$ (nm) is much smaller than the state-of-the-art light intensity of fluorescence microscopes. $S_m$ is assumed to change between low and high levels of strain, i.e., from $0.005\%$ to $1\%$, with higher ultrasound pressures for cellular tracking. For the MPT simulation in Section \ref{s6s2}, SNR is defined with respect to the average $(\mu_1 \, + \, \mu_2)   \, / \, 2$ and chosen between  $-7$ to $10$ dB. Next, FM modulation with the multiplier is numerically simulated, and then the radar tracking system for microfluidic applications is simulated with Monte-Carlo simulation experiments.
  
\subsection{Frequency Multiplier and Optical FM Modulation}
\label{s6s1}

\begin{figure*}[!t]
\centering
\includegraphics[width=2.35in]{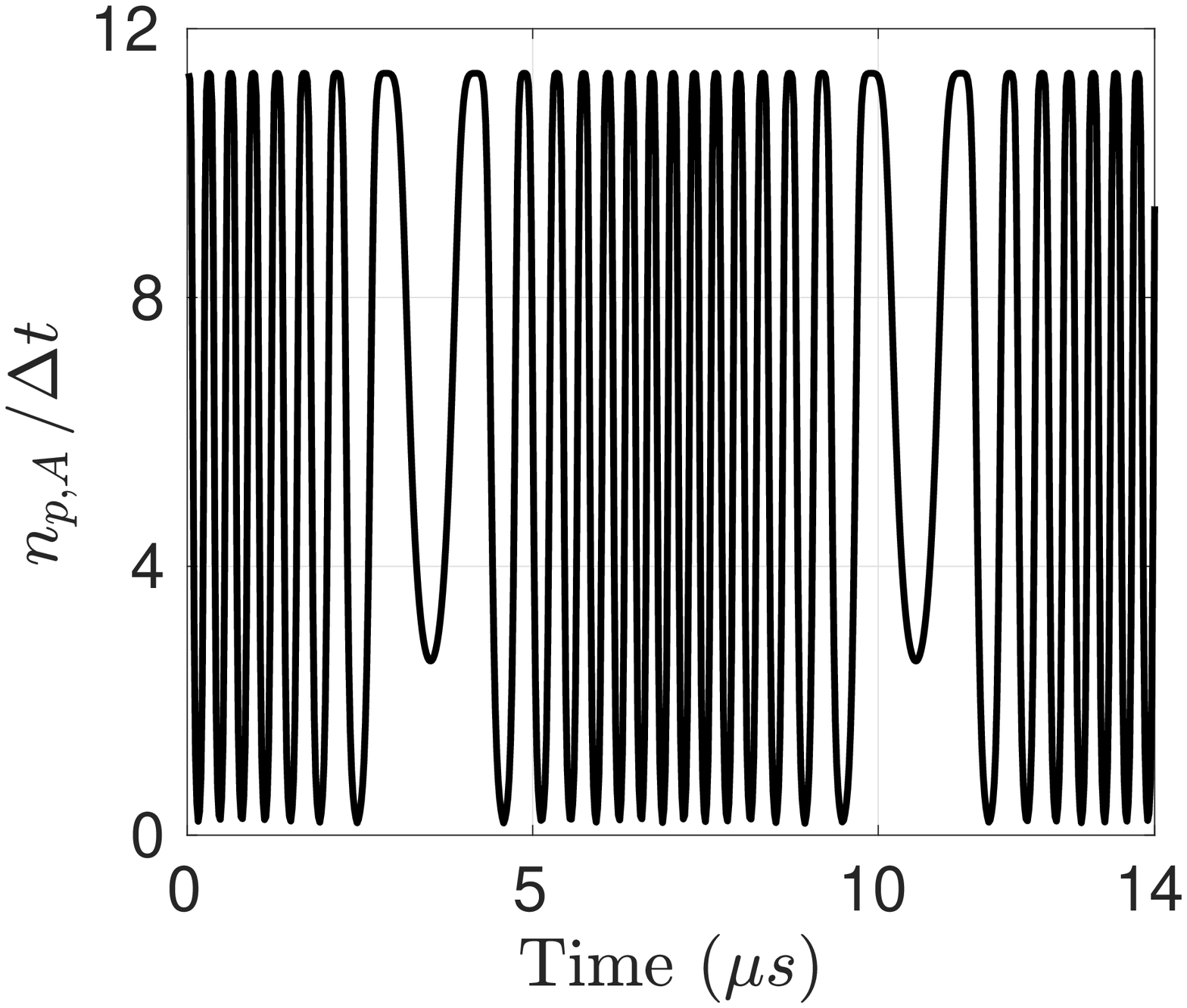} \includegraphics[width=2.35in]{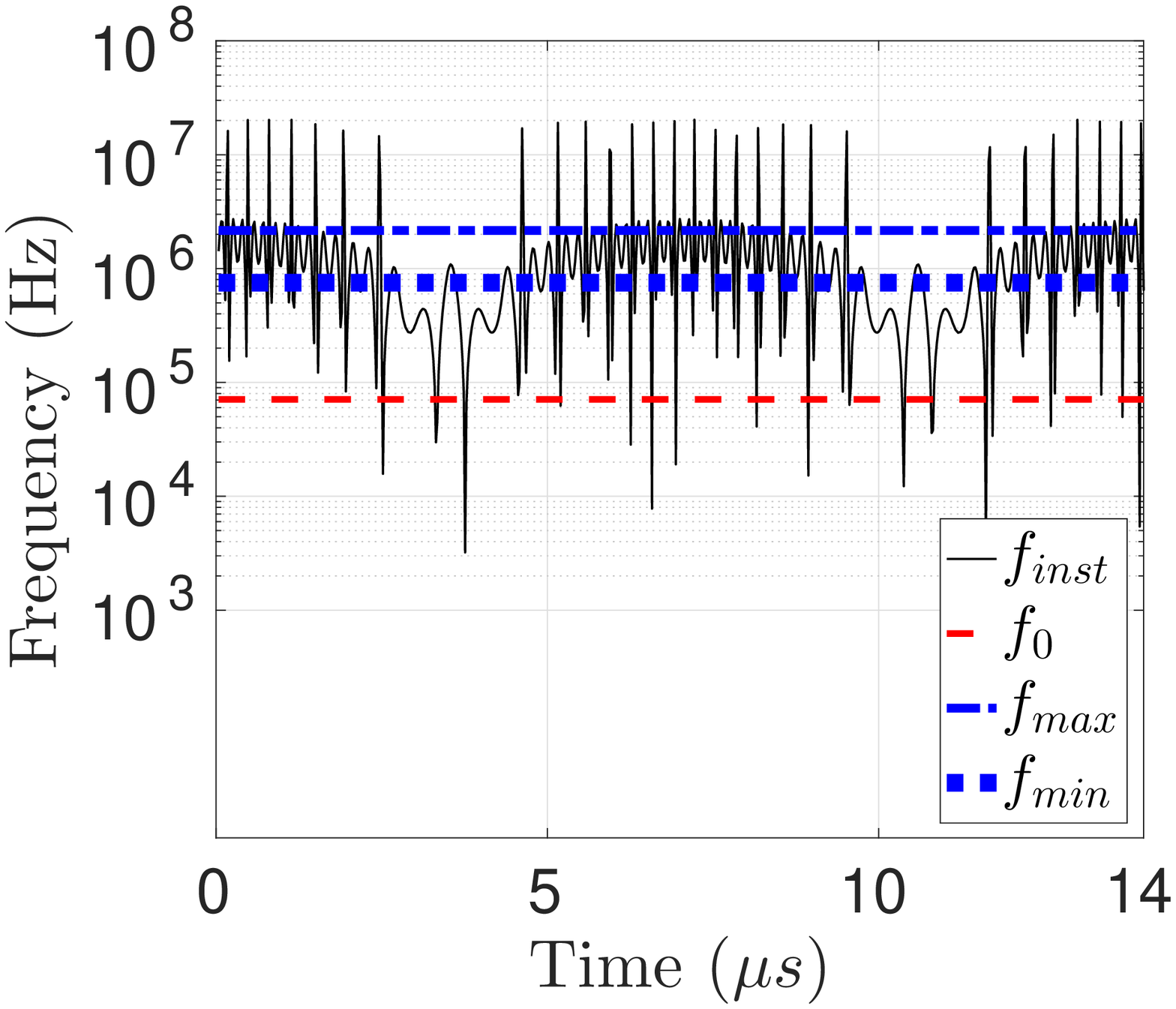} \includegraphics[width=2.35in]{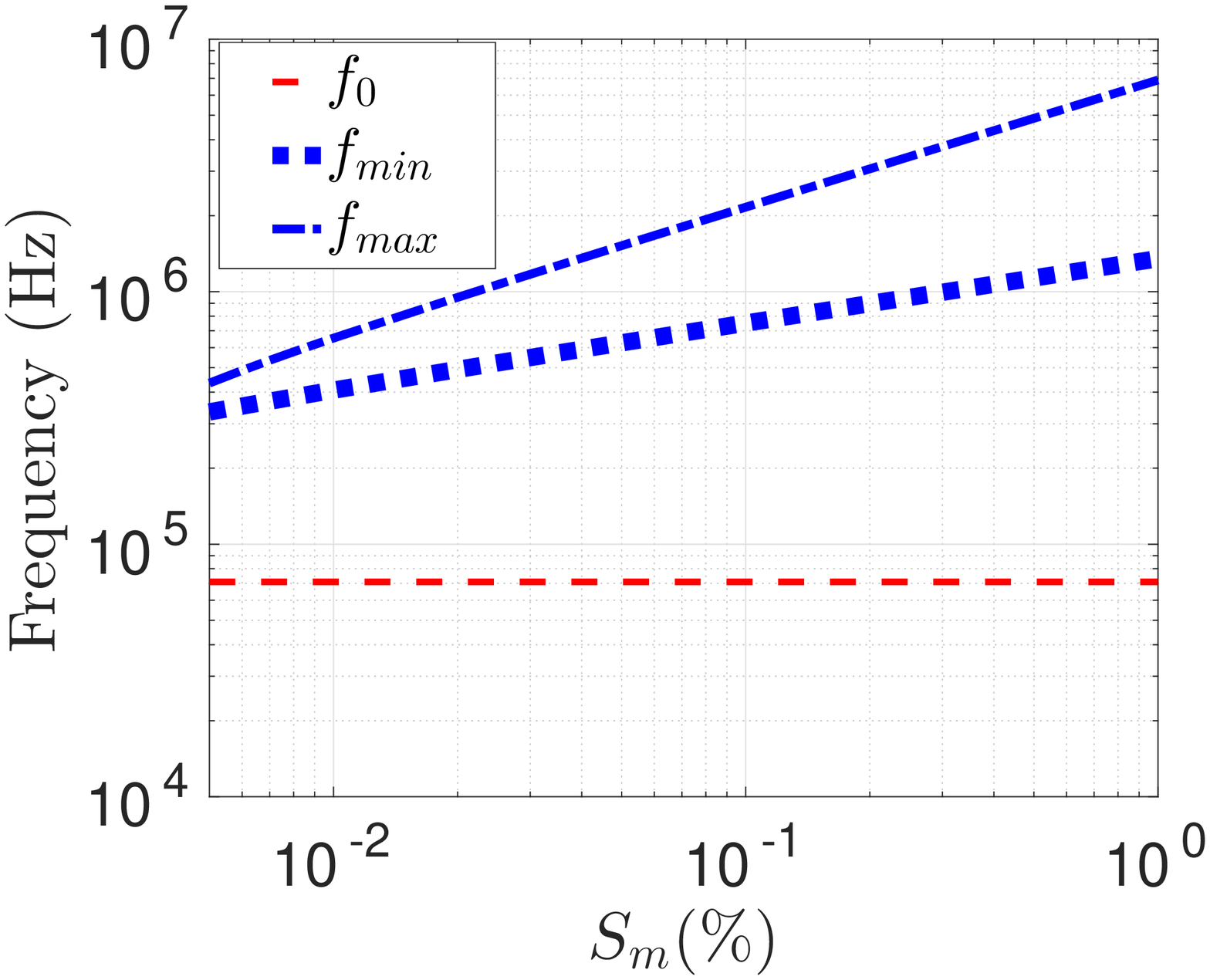} \\
(a) \hspace{2.2in}  (b)  \hspace{2.2in} (c)
\caption{(a) Acceptor emission photon count $n_{p, A}(i \Delta t)$ and  (b) the instantaneous frequency by Hilbert transform for varying time intervals in a single acoustic period with $f_0  \approx 71$ KHz, $\Delta t = 20$ ns, $k_{\Delta  A} =   k_{\Delta  D} = 4$, $L_g = 10 \, \mu$m and $S_m = 0.1\%$,  and (c) $f_{min}$-$f_{max}$ of the instantaneous frequency for varying $S_m$ from $0.005\%$ to $1\%$.}
\label{figtag1abc}
\end{figure*}

FM modulation and the frequency multiplication effect are simulated for periodically placed D-A layers as shown in Fig. \ref{figMult}  while assuming  $k_{\Delta A} = k_{\Delta D} = 4$ and $\Delta t = 20$ ns as two times of the total average time for photon emissions  in \cite{qd1} validating simulations  reaching $f_{max} = 50$ MHz. This is larger than the simulated frequencies as shown in Fig. \ref{figtag1abc}(c).   The total number of D-A pairs is calculated as $\approx 12 \times 10^6$ for the device dimensions of $10 \, \mu$m $\times$    $10 \, \mu$m $\times$    $10 \, \mu$m  and the weight of approximately $23$ picograms compared with average weight of the human cell on the orders of nanograms while assuming the acceptors reside on carbon membranes of the thickness of one nanometer, the width of $2 \, R_A$ and the length of $L_g \, / \, 50$ with the same mass density of the graphene as shown in Fig. \ref{figMult}.  In Fig. \ref{figtag1abc}(a), the acceptor emission photon count  $n_{p, A}$ per $\Delta t$ is shown while the corresponding   $f_{inst}$, $f_{min}$ and $f_{max}$ are shown in Fig. \ref{figtag1abc}(b). The frequency multiplication and FM modulation with instantaneous frequency changes as described in Fig. \ref{figtag2} are clearly observed for the device designed with respect to $S_m = 1\%$ (to determine $N_{l,A}$ as discussed in Section \ref{s3s1})  while operating at $S_m = 0.1\%$.  In Fig. \ref{figtag1abc}(c), the increase in $S_m$ results in faster  velocities for the donors passing through layers and with higher frequency multiplication effect. Next, the proposed architecture is modified with special layer arrangements to realize different CSS sequences.

\subsection{Single and Multiple Particle Tracking}
\label{s6s2} 

\begin{figure}[!t]
\centering
\includegraphics[width=3.5in]{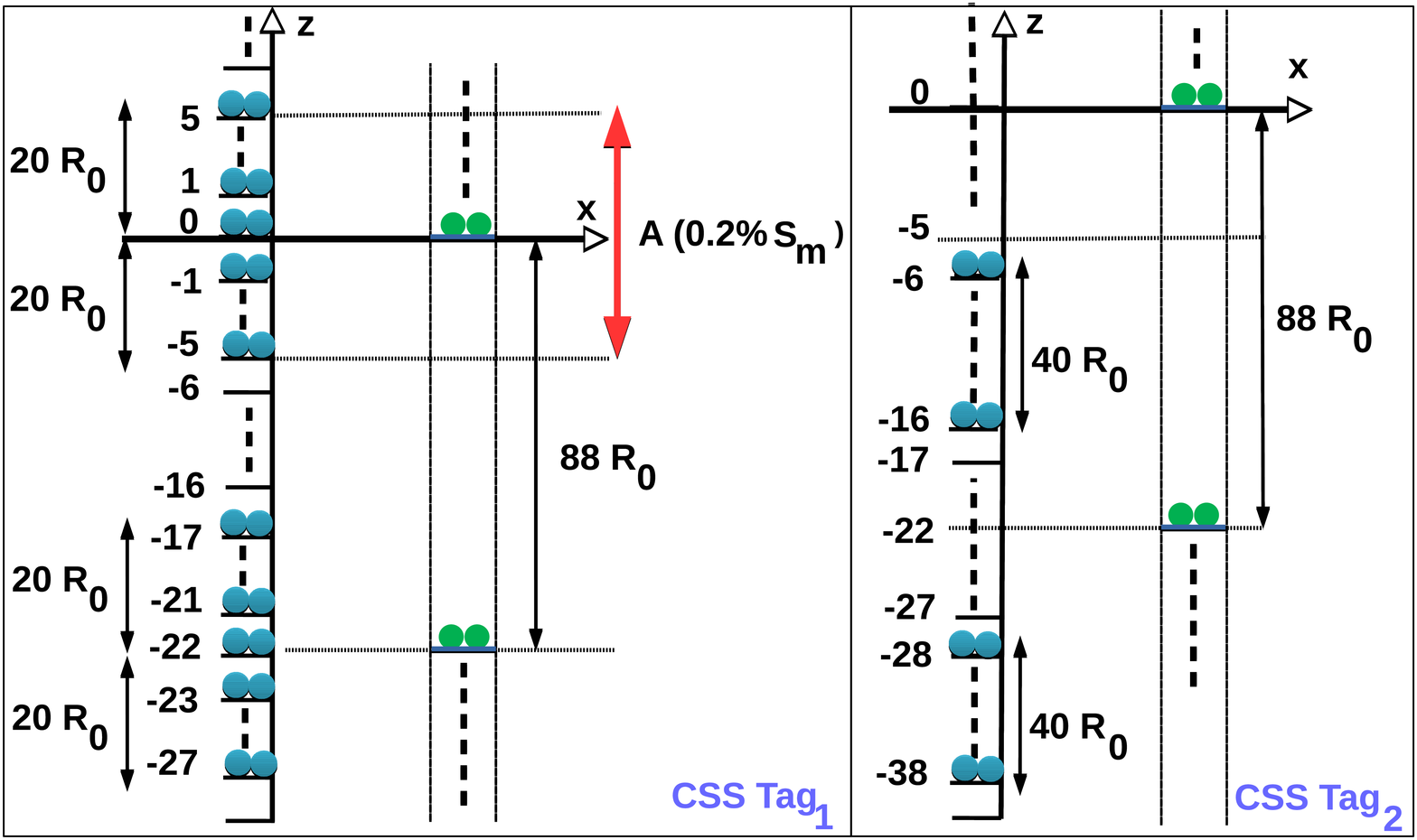}  
\caption{Two basic device structures for different CSS tags where $Tag_1$ and $ Tag_2$ result  in higher and lower frequency chirps, respectively, designed with respect to $S_m \, < \, 0.2 \%$, and with separated groups of acceptor layers having inter-layer distance parameter $k_{\Delta A} = 4$  while individual donor layers are separated by $k_{\Delta D}\, = 22 \, k_{\Delta A}$. }
\label{figtag4}
\end{figure} 

\begin{figure}[!t]
\centering
\includegraphics[width=2.5in]{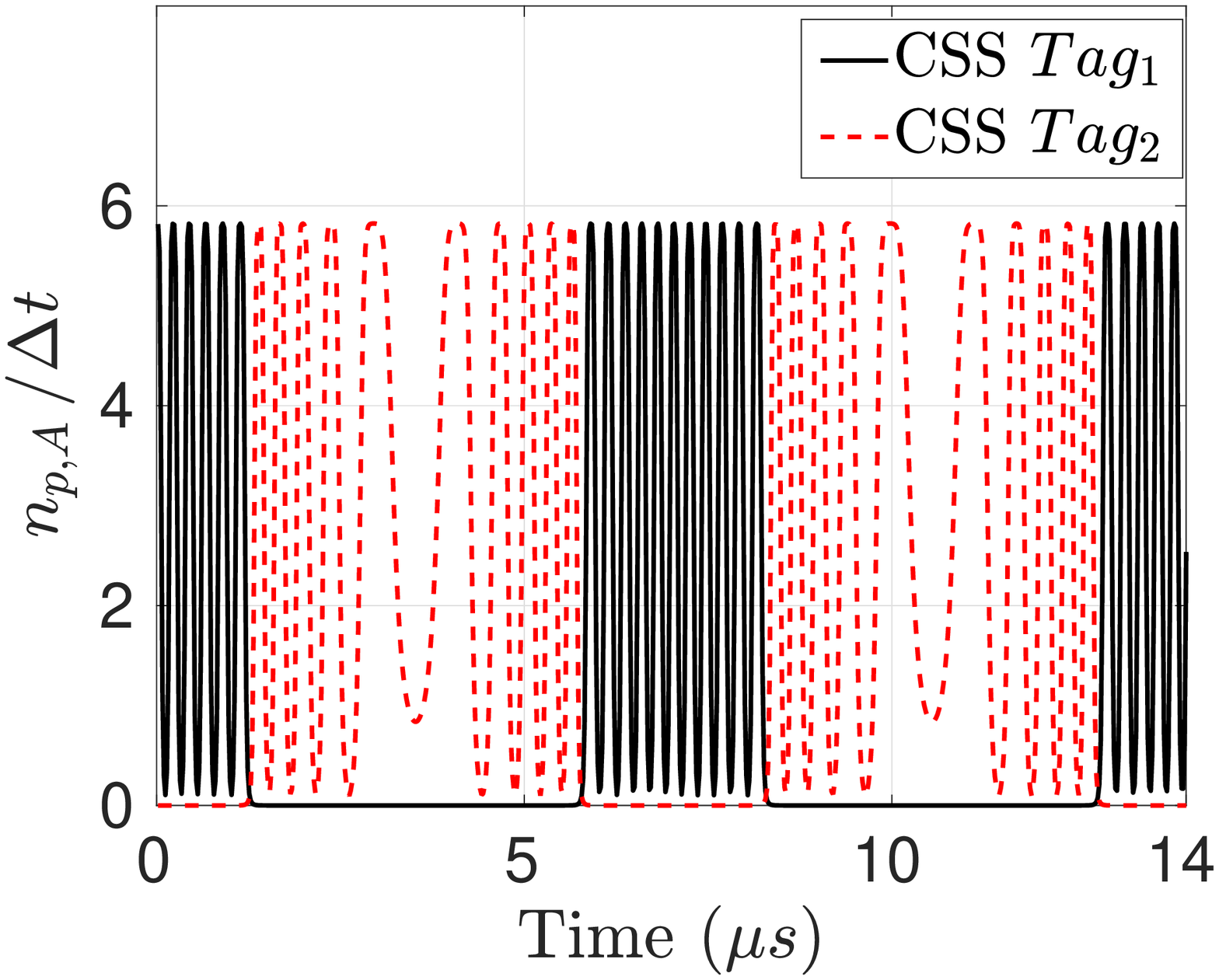} \\
\hspace{0.2in} (a) \\
\includegraphics[width=2.5in]{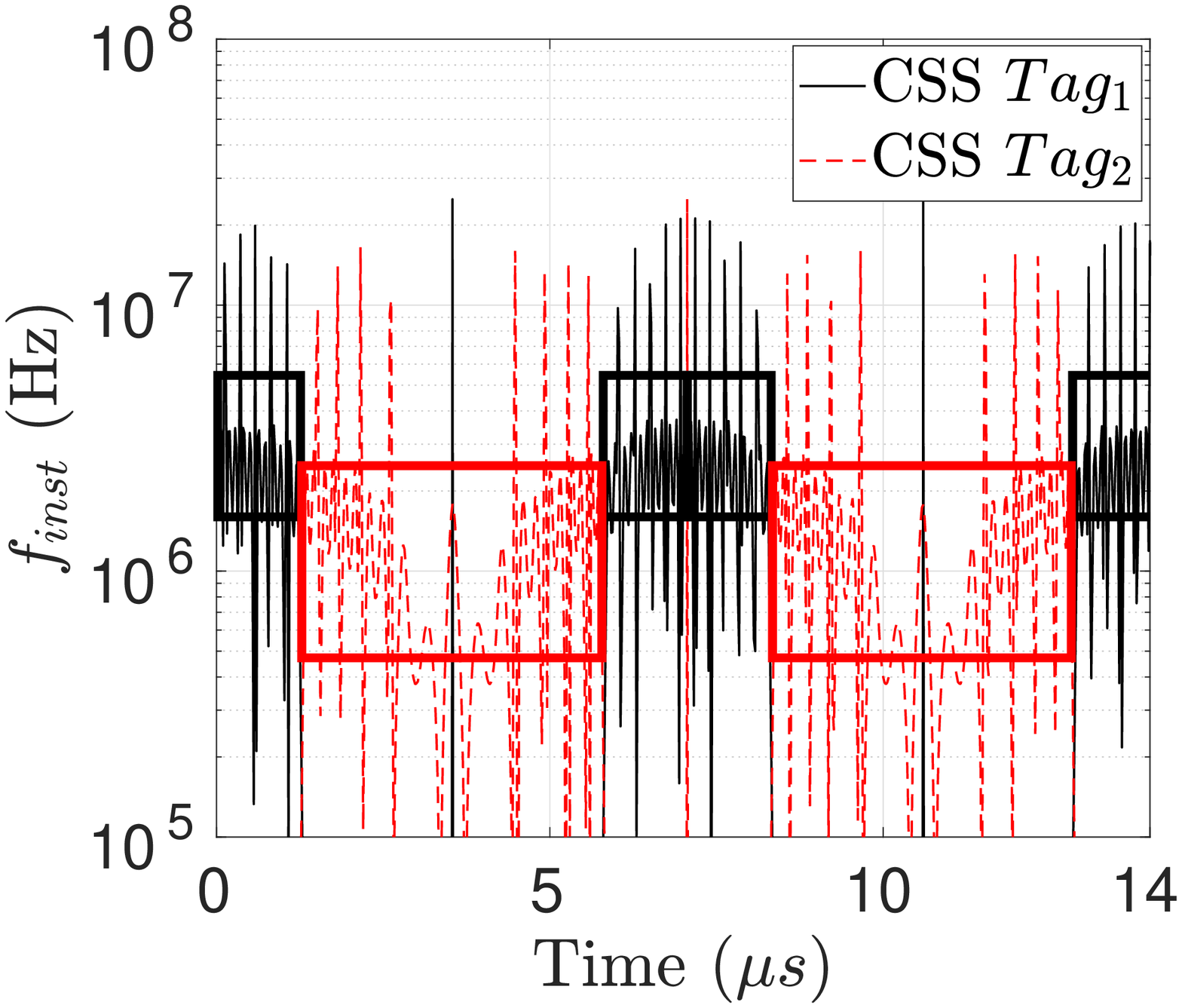} \\
\hspace{0.2in}  (b) 
\caption{(a) Time domain waveforms and (b) the instantaneous frequencies of the $Tag_1$ and $Tag_2$ CSS sequences.}
\label{figtag56}
\end{figure} 

\begin{figure*}[!t]
\centering
\includegraphics[width=2.35in]{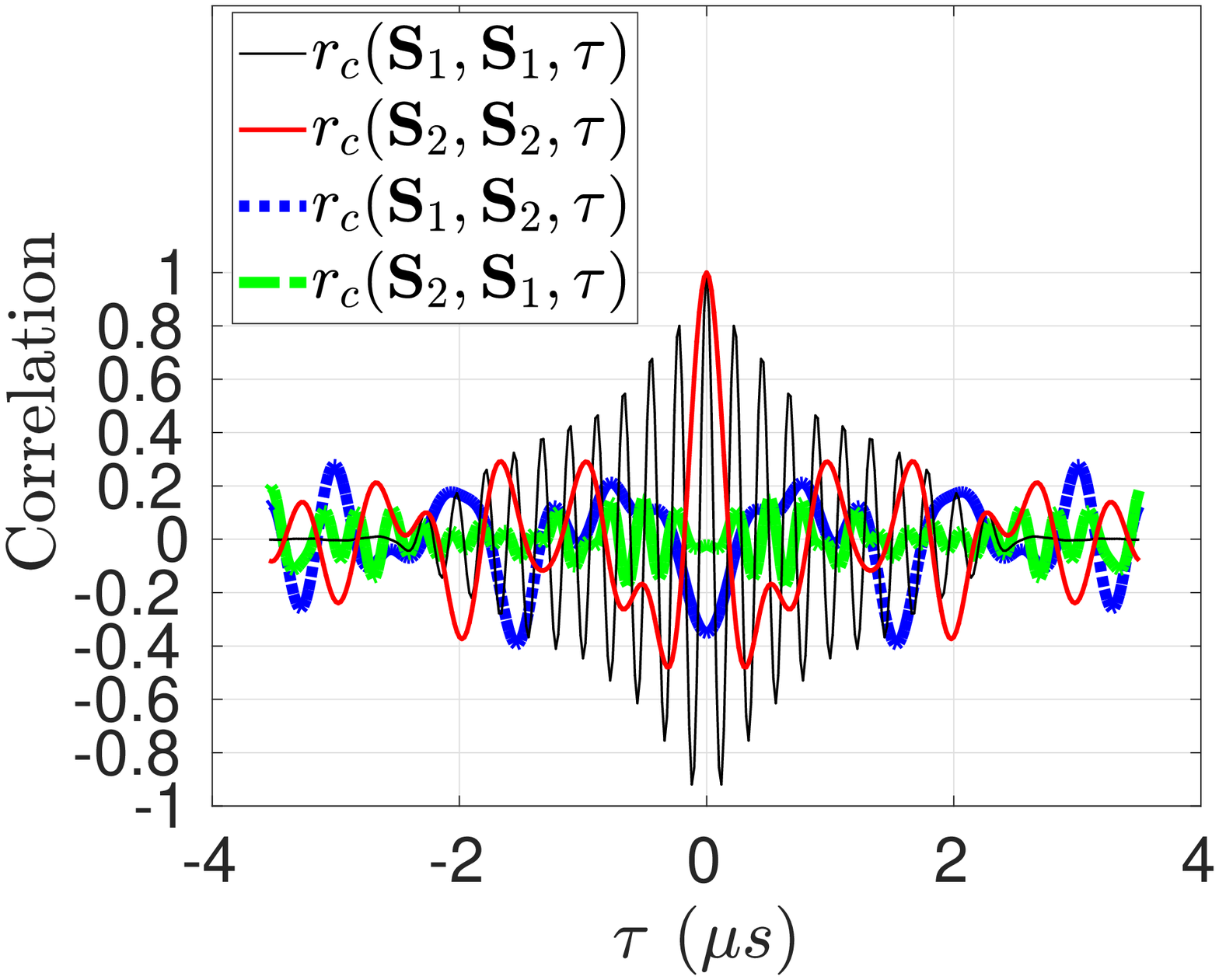} \includegraphics[width=2.35in]{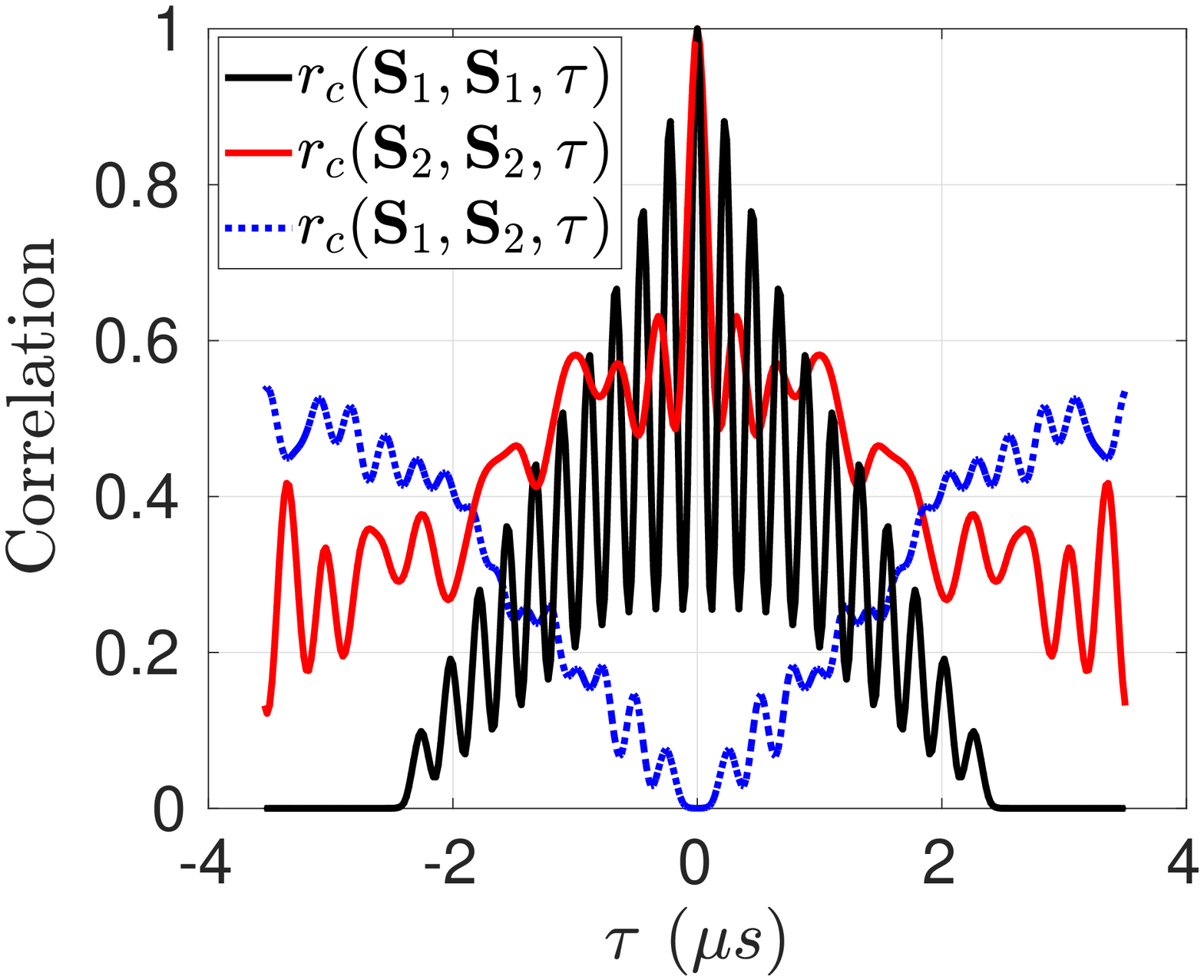}  \includegraphics[width=2.35in]{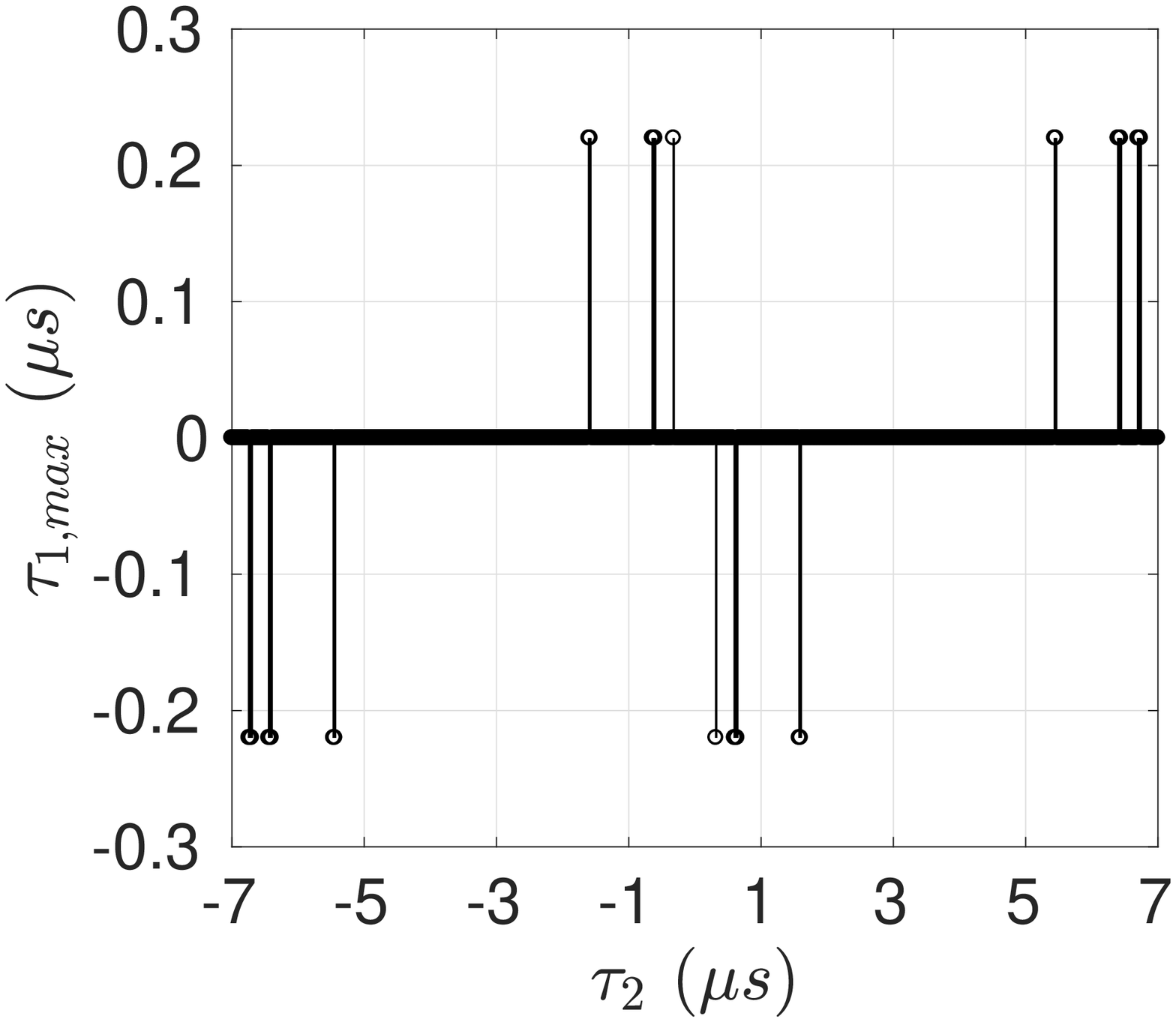} \\
\hspace{0.1in}  (a) \hspace{2.1in}  (b) \hspace{2.1in}  (c)
\caption{The correlations for $Tag_1$ and $Tag_2$ CSS sequences (a) with and (b) without TF filtering, and (c) the effect of $Tag_2$ interference on $Tag_1$ shifting the maximum autocorrelation point of $Tag_1$ by $\tau_{1,max}$ for varying $\tau_2$ in $r_c(\mathbf{S_{1}(t)}  + \mathbf{S_{2}}(t-\tau_2),\mathbf{S_{1}(t)}, \tau_1)$.}
\label{figtag7ab}
\end{figure*} 

The mechanical structure of the specially designed tags is shown in Figure \ref{figtag4} realizing two different CSS sequences, i.e., $Tag_1$ and $Tag_2$. 
The device is designed with $S_m \, < \, 0.2\%$ tunable with respect to the application. The distance between the donor layers is increased with $k_{\Delta D}\, = 22 \, k_{\Delta A}$ while $k_{\Delta A}$ is kept at four. The same device dimension described in Table \ref{table2} results in a modulator with the total number of donor and acceptor molecules of $\approx 6 \times 10^5$ and $\approx 6.5 \times 10^6$, respectively, and the total weight of $\approx 3$ picograms.  The groups of acceptor layers are formed by shifting their relative positions with respect to the donor layers to form different time and frequency bands in the CSS sequences $Tag_1$ and $Tag_2$ as shown in Fig. \ref{figtag56}(a) and (b). The rectangular boxes in Fig. \ref{figtag56}(b) show the TF support of the two tags where TF filtering synchronized with the carrier period separates two tags in the receiver. 

\begin{figure}[!t]
\centering
\includegraphics[width=2.7in]{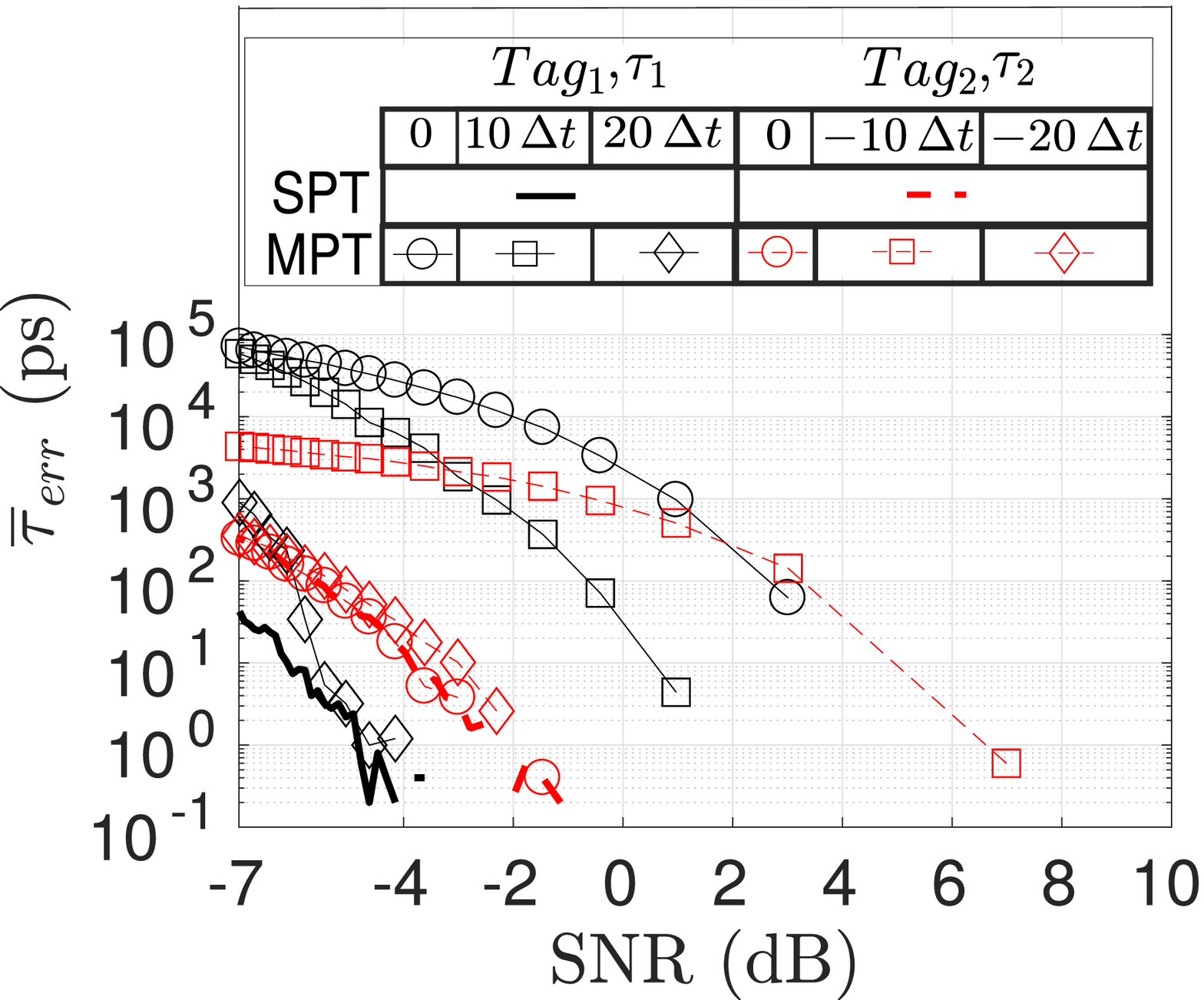}  
\caption{ Average estimation errors for the delays of $Tag_1$ and $Tag_2$, i.e., $\tau_1$ and $\tau_2$, for varying SNR for SPT and MPT  where $\tau_2 = -\tau_1$ for MPT.}
\label{figtag9}
\end{figure}

In Fig. \ref{figtag7ab}(a), the auto and cross correlations between the tags are shown where the auto-correlations show oscillating  and decaying correlations while cross-correlations are smaller than the auto-correlations with oscillating magnitude. The envelope of $Tag_1$ decays more smoothly with a lower performance with respect to $Tag_2$. Since the effect of Doppler velocity is not taken into account, the optimum receiver performance defined in Fig. \ref{figtag9} is experimented by excluding the filtering steps in the remaining simulations. In Fig. \ref{figtag7ab}(c),  the effect of the interference of $Tag_2$ on $Tag_1$ is simulated showing the peak point   $\tau_{1, max}$  of  $r_c(\mathbf{S_{1}(t)}  + \mathbf{S_{2}}(t-\tau_2),\mathbf{S_{1}(t)}, \tau_1)$ with respect to $\tau_1$ for varying $\tau_2$. It is observed that the peak of $Tag_1$ jumps from the maximum autocorrelation point to the neighbor peaks due to interference from $Tag_2$ for a finite set of $\tau_2$ delays. The same performance degradation occurs for $Tag_1$ in the high noise regime similarly as shown in the following numerical simulations in Fig. \ref{figtag9}.  It is an open issue to design better codes with  orthogonal and fast decaying correlations. In simulations, it is assumed that the particle distances  are not coinciding with the critical delay points.

The total emission optical power for the proposed sequences in Fig. \ref{figtag56}(a) with the mean photon count of $\mu_{av} = (\mu_{1} \, + \, \mu_{2}) \, / \, 2$ is calculated as approximately $P_{e} = \mu_{av} \, h_{p} \, v_f  \, / \, \Delta t =  2$ nW. A basic photodetector of the responsivity $R_p = 1 \, A/W$ and the idealistic case with the full photon collection efficiency result  in shot noise power $2 \, q \, \Delta f \, R_p \, P_{e} \approx 2.5 \, 10^{-21}$ ($Amp^2$) for the bandwidth of $\Delta f \approx 5$ MHz  shown in Fig. \ref{figtag56}(b) and the thermal noise power of $4 \, \kappa_B \, T \, \Delta f  \, / \, R_{eq} \approx 8.2 \, 10^{-21}$ ($Amp^2$) with the assumption $R_{eq} = 10$ M$\Omega$ where $\kappa_B  \, = \, 1.38 \times 10^{-23}$ J/K is the Boltzmann constant and $T\, = \,  300$ K is the room temperature \cite{bg3}. The signal power is approximately $4 \, 10^{-18}$ ($Amp^2$) which is much larger than the noise power. The SNR can be increased with PMTs significantly. 

In Fig. \ref{figtag9}, the Monte-Carlo simulation results for varying SNR of $r_T(t)$ are shown where  $\tau_1$ and $\tau_2$ are the delays corresponding to the vertical positions of $Tag_1$ and $Tag_2$ chosen in the range $\pm [0, 5, 10] \times \Delta t$ corresponding to the vertical distances with respect to reference modulator approximately in $[-155, 155] \, \mu$m computed with $v_e \times \tau $.  The number of simulation experiments is set to $ 10^5$ while the average estimation errors denoted by $\overline{\tau}_{err}$ for $\tau_1$ and $\tau_2$ are shown for SPT cases where only one particle is present in the channel and MPT case  where both the particles are present. As SNR approaches $\approx -4$ dB and $\approx -1$ dB for SPT of $Tag_1$ and $Tag_2$, respectively, the average estimation errors  are shown to decrease to zero providing a highly error tolerant SPT mechanism. Besides that, the average error on the orders of tens of picoseconds (ps) for the simulated levels of high noise is lower than the maximum resolution $\Delta t  = 20$ ns corresponding to  $v_e \times \Delta t \approx 30 \, \mu$m resolution in the z-axis. Therefore, even low levels of SNR values allow accurate cell positioning due to the spreading of the CSS sequences. On the other hand, in MPT case, the simulations are realized for $\tau_1 = - \tau_2$ and the estimation error decreases as  the distance between the particles increases with lower ISI. The performance of  $Tag_{1}$ is worse than $Tag_2$ due to slower decaying envelope. The estimation errors decrease to zero for SNR values approximately larger than $7$ dB providing an accurate MPT mechanism. The design of CSS sequences, and the performance with respect to the effects of the channel and Doppler shift with high velocity particles are open issues.

\section{Conclusion}
\label{conclusion}

In this article,  a novel nanoscale acousto-optic radar architecture and particle tracking system is presented which significantly improves the state-of-the-art fluorescence based imaging systems with signaling based tracking capability. The system is based on the nanoscale optical modulator structures utilizing  VFRET and vibrating CdSe/ZnS QDs on graphene membranes. The proposed modulator achieves acousto-optic frequency multiplication, generates CSS sequences having different TF supports and exploits the unique properties of graphene and  QDs. The optical CSS sequence generation mechanisms are simulated numerically with modulators having micrometer dimensions and weights of several picograms, and low level of light intensities compared with classical laser excitations. Monte-Carlo simulations are realized for two particle tracking experiment showing significant performance for MPT tasks with SNRs in the range $-7$ dB to $10$ dB and  high speed tracking capability. CSSTag system can be utilized in all the fluorescence based imaging applications enhancing their capabilities such as diagnostic microfluidic and in-body applications of cell identification, cytometry and cellular imaging. 

\section*{Acknowledgment}
This work is supported by Vestel Electronics Inc., Manisa, 45030, Turkey.

\vspace{-1.5cm}
%
 
\vspace{-1.5cm}
%

\end{document}